\documentclass[fleqn,10pt]{wlscirep}
\usepackage[utf8]{inputenc}
\usepackage[T1]{fontenc}
\usepackage{lineno}
\usepackage{xcolor}
\usepackage{multirow}
\usepackage{hyperref}
\linenumbers

\newcommand{\exciting}{{\usefont{T1}{lmtt}{b}{n}exciting}}

\newcommand{\ie}{i.e., }
\newcommand{\eg}{e.g., }

\title{Extrapolation to complete basis-set limit in density-functional theory by quantile random-forest models}

\author[1, 2]{Daniel T. Speckhard}
\author[2]{Christian Carbogno}
\author[1, 2]{Luca Ghiringhelli}
\author[1]{Sven Lubeck}
\author[2]{Matthias Scheffler}
\author[1]{Claudia Draxl}

\affil[1]{Physics Department and IRIS Adlershof, Humboldt-Universit\"at zu Berlin, Zum Gro\ss en Windkanal 2, Berlin, 12489, Germany}
\affil[2]{NOMAD Laboratory at the Fritz Haber Institute of the Max Planck Society and IRIS Adlershof, Faradayweg 4-6, Berlin, 14195, Germany}

\affil[*]{corresponding author(s): Daniel Speckhard (speckhard@fhi.mpg.de)}

\begin{abstract}
The numerical precision of density-functional-theory (DFT) calculations depends on a variety of computational parameters, one of the most critical being the basis-set size. The ultimate precision is reached with an infinitely large basis set, \ie in the limit of a complete basis set (CBS). Our aim in this work is to find a machine-learning model that extrapolates finite basis-size calculations to the CBS limit. We start with a data set of 63 binary solids investigated with two all-electron DFT codes, \exciting\ and FHI-aims, which employ very different types of basis sets. A quantile-random-forest model is used to estimate the total-energy correction with respect to a fully converged calculation as a function of the basis-set size. The random-forest model achieves a symmetric mean absolute percentage error of lower than 25\% for both codes and outperforms previous approaches in the literature. Our approach also provides prediction intervals, which quantify the uncertainty of the models' predictions.
\end{abstract}
\begin{document}
\nolinenumbers
\flushbottom
\maketitle

\thispagestyle{empty}


\section*{Introduction}
The assessment of the quality of density-functional-theory (DFT) calculations concerns the accuracy of the exchange-correlation functional and the numerical precision that depends on a variety of computational parameters. This paper deals with the latter, of which a most critical parameter is the size of the basis set. Only with an in-principle infinitely large basis-set size, the result of the calculation is as precise as possible for the chosen exchange-correlation functional. This limit is known as the complete basis-set (CBS) limit \cite{hill2009extrapolating}. However, a basis-set size approaching this limit, would take infinite time to compute. Therefore, in practice, the basis set is truncated at a size that balances precision and computational cost. Extrapolation from low-precision settings to the CBS limit is commonplace in quantum chemistry \cite{bakowies2007extrapolation}. In materials science, convergence tests with respect to the basis-set size are typically done, but extrapolation to the CBS limit is not often performed. Our aim in this work, is to find a model that can extrapolate the result of a DFT calculation to the CBS limit. More specifically, we seek to predict the difference in the total energy per atom computed with an incomplete basis-set size to a computation performed in the CBS limit. We exemplify our approach with a data set of binary materials. Note, the extrapolation to the CBS limit depends on the chosen functional. Here, we only consider the PBE functional of the generalized-grandient approximation (GGA).

Our motivation is two-fold. First, such a model opens up the possibility to perform less precise, and therefore computationally less demanding {\it ab initio} calculations to predict a more precise result. Here, we recall that typical DFT-GGA implementations scale with order $\mathcal{O}(N^3)$ where $N$ is the basis-set size. Second, it allows us to assign uncertainty estimates to the huge amounts of {\it ab initio} data contained in open-access databases. For instance, the Novel Materials Discovery (NOMAD) Respository \cite{draxl2019nomad} currently hosts about 140 million ground-state DFT calculations. These calculations were carried out for a variety of purposes ranging from molecular-dynamics simulations of complex systems with less precise settings to ultra high-precision calculations for elemental solids. Uncertainty estimates for the total energies would provide users useful information about the precision of these calculations and how the data can be re-used/re-purposed~\cite{andersen2021optimade}. If one can do CBS extrapolation, one could even extrapolate all these data to the CBS limit, which would be even more useful.

In this work, we train a quantile-random-forest (QRF) model to predict the total-energy difference $\Delta E^{AB}$ for binary materials containing the two elements \emph{A} and \emph{B}. We train on a data set consisting of DFT results for 71 elemental and 63 binary solids, computed by the two full-potential all-electron codes FHI-aims and \exciting\ with varying basis-set size. For details concerning the data set we refer to Ref. \citeonline{carbogno2022numerical}. The linearized augmented planewave (LAPW) code \exciting\ employs augmented planewaves (APW) plus local orbitals (LO) as its basis, while FHI-aims uses numeric atom-centered orbitals (NAOs). These two codes are representatives of all-electron, full potential packages. This means that they can simulate the behavior of all electrons in a material on the same footing and are proven to be among the most precise DFT codes available~\cite{lejaeghere2016reproducibility}. Despite their significantly different concepts, algorithms, and numerical approaches, both codes are expected to give close-to identical results~\cite{lejaeghere2016reproducibility}. Due to the very different basis sets and thus numerical implementations, we investigate the behavior of their convergence separately. We compare our modeling efforts to a stoichiometric model which was introduced in Ref. \citeonline{carbogno2022numerical} and find that our models outperform the latter in terms of several important metrics. 

\section*{Methods}

We formulate our task of extrapolation to the CBS limit as a $\Delta$-learning problem~\cite{bogojeski2020quantum}. We are given the results of a single DFT calculation at a fixed basis-set size, $N_b$, which is smaller than that of the converged case, $N_\infty$, known as the CBS limit. The data set in Ref. \citeonline{carbogno2022numerical} defines the total-energy convergence criteria with respect to the basis-set size as $10^{-4}$ eV/atom. The data is fed into a statistical learning algorithm to estimate the difference between the imprecise DFT calculation and the CBS limit. Our task, in other words, is to estimate the total energy per atom of a binary material composed of elements \emph{A} and \emph{B}, in the CBS limit, $E^{AB}(N_\infty)$, using the results of a DFT calculation with the fixed incomplete basis-set size, $N_b$. Mathematically, we aim at finding the change ($\Delta$) in total-energy from an incomplete basis-set size, $E^{AB}(N_b)$, to the CBS limit $E^{AB}(N_\infty)$. As can be seen in eq. \ref{eq:delta_learning}, we target $\Delta E^{AB}(N_b)$.
\begin{equation}
E^{AB}(N_\infty) = E^{AB}(N_b) + \Delta E^{AB}(N_b)
\label{eq:delta_learning}
\end{equation}
We employ QRF models for obtaining the total-energy differences $\Delta E^{AB}$ per atom. Other DFT settings are kept constant. Physically, this means we aim to use an imprecise, less computationally intense, calculation that gives us $E^{AB}(N_b)$ in tandem with a statistically learned model that predicts $\Delta E^{AB}(N_b)$. Together with these two sources, the imprecise calculation and model, we predict the total-energy of the complete basis-set limit.

\subsection*{Stoichiometric Model}
Our baseline model, to compare our new approach with, is a stoichiometric model introduced in Ref. \citeonline{carbogno2022numerical},
\begin{equation}
\begin{split}
\label{eq:stoichiometric_equation}
\Delta E^{AB}(N_b) = C^{A}*\Delta E^{A}(N_b) + C^{B}*\Delta E^{B}(N_b) .
\end{split}
\end{equation}
Each binary solid, represented as \emph{AB}, is composed of two chemical elements, labeled \emph{A} and \emph{B}. Here, the letter \emph{A} (\emph{B}) refers to the less (more) electronegative element in the binary. $\Delta E^{A}(N_b)$ refers to the CBS total-energy correction for the corresponding lowest-energy elemental solid of element $A$ when using a basis-set size, $N_b$. $C^{A}$ is the stoichiometric fraction that the element $A$ appears in the binary.

\subsection*{Basis Set of \exciting}
\label{sec:basis_set_exciting}
The most important parameter determining the quality of augmented plane-wave basis sets is $RK_{\text{max}}$, which is the product of the radius of the smallest atomic (muffin-tin) sphere and the plane-wave cutoff. In Ref.~\citeonline{carbogno2022numerical}, a {\it precision factor}, $(RK_{\text{max}}/RK_{\text{max}}^{\text{opt}})^2$ was introduced which captures the precision of the basis set of \exciting\ quite well. The data set we use contains elemental solids and binaries at the same percentage value of the precision factor. However, $RK_{\text{max}}^{\text{opt}}$ may be different for a binary and its elemental solids. Note, in this data set, the number of  APWs is varied but the number of LOs is kept constant. More information about the basis-set precision parameter for \exciting\ is given in the supplementary information of Ref.~\citeonline{carbogno2022numerical}.

\subsection*{Basis Set of FHI-aims}

FHI-aims offers tabulated species-specific suggestions for numerical settings and NAOs, named as ``light'', ``tight'', or ``really tight'' defaults. In general, one does not need to use the tabulated settings. However, in this work we do. On top of the numerical settings defined in these defaults, we also consider different "basis-set size settings" (minimal, standard, tier1, or tier2). The combination of both ultimately dictates the number and type ($s$, $p$, $d$, etc.) of basis functions included in a calculation. "Standard" refers to the default basis-set size suggested in the respective numerical setting. The difference in the number of NAOs per valence electron from the CBS limit, labeled $\Delta SB^{AB}_{PVE}$, is used as a basis-set size metric. More information about the basis-set size in FHI-aims is given in the supplementary information of Ref.~\citeonline{carbogno2022numerical}.

\subsection*{Model Features}

We feed the QRF model information about the two elements in terms of elemental solids by providing $C^A*\Delta E^{A}(N_b)$ and $C^B*\Delta E^{B}(N_b)$ which are the two terms in the stoichiometric model from eq.~\ref{eq:stoichiometric_equation}. Atomic information about the elements in terms of isolated atoms is also provided. Their use is motivated by other statistical learning models in materials science~\cite{ghiringhelli2015big}. The electron affinity ($EA^A$, $EA^B$), the ionization energy ($EI^A$, $EI^B$) and the mean radius ($r_{s}^{A}$, $r_{s}^{B}$) for the $s$-like pseudo orbital of element $A$/$B$ computed with FHI-aims are fed into the QRF models.

\subsection*{Quantile Random Forests}
We choose to use random-forests (RF) based methods for CBS extrapolation since RFs are known to perform well on a wide range of tasks and require minimal tuning and no data scaling~\cite{ali2012random}. We provide here a brief summary of the quantile random-forest method introduced by Meinshausen \cite{meinshausen2006quantile}. Random forests are a collection of decision tree models. A decision tree is a piece-wise constant model. Decision trees work by partitioning the input feature space into discrete regions. The tree then assigns a constant estimate for that region depending on the training data points that fall in that region~\cite{hastie2009elements}. The decision tree chooses the splits (and therefore the discrete regions of the input data space, using a greedy algorithm. This means each new split minimizes the optimization metric for that split and not for potential subsequent splits. In order to make a prediction on an input data point, the decision tree looks at the region in which the input data point falls. It then predicts the constant prediction for that region node.

Random forests are built by sampling the training data with replacement (bootstrapping) and fitting separate decision trees that are forced to use only a random subset of model features. This randomness forces the many decision trees in the random forest to have different splits. The random forest model suffers less from overfitting than a single decision tree~\cite{breiman2001random}. To make an inference, the random forest looks at what region the input data point fall into for each tree. Each tree therefore has an assigned constant for the input data point. Since there are several trees in the forest, a data point falls into a leaf node for each tree, and the random forest predicts the average of the constants from each tree.

QRFs further examine the leaf nodes. When performing predictions, QRFs look at the leaf node into which the input data point falls, for each tree in the forest. The QRF creates quantiles (\eg $2.5\%$ and $97.5\%$) by sorting all of the inferences that each tree in the forest predicts for that data point. These quantiles are used as statistically meaningful prediction intervals. The median quantile can be used for inference. In this work, we continue to use the mean of the constant estimates (which is typical for a RF) for inference and use the QRF for the prediction intervals.

In this work, we apply the QRF method in a regression setting to minimize the root-mean-squared-logarithmic-error (RMSLE) metric. We derive, in the supplementary information section, how decision-tree parameters are learned when optimizing for the RMSLE metric. Our data is randomly split into training and test data using an 80/20 split. We perform ten-fold grid-search cross-validation (CV) to choose the number of decision trees that comprise the random forest, the minimum number of samples per leaf, and the fraction of features considered at each split. More details on the cross-validation can be found in the accompanying Jupyter notebook.

\subsection*{Combination Model}

We also investigate the performance of a QRF model trained on the residuals (remaining error) of the stoichiometric model. In mathematical terms, the residual to which we fit is the variable: $
\Delta E^{AB}(N_b) - C^{A}*\Delta E^{A}(N_b) - C^{B}*\Delta E^{B}(N_b)$. This means our estimate of the CBS total-energy correction is the sum of the stoichiometric model and a new QRF trained on the stoichiometric residuals. We call this model the "combination model". The motivation is that subtracting the stoichiometric model from the target might allow the QRF to focus more on non-linear contributions.

\subsection*{Metrics}

Our DFT data used for the training contain CBS energy corrections ranging in magnitude from $10^{-6}$ eV/atom to several eV/atom. The root-mean-squared-error (RMSE) and mean-absolute-error (MAE), which are common loss functions, are known to give more weight to large targets in a data set that spans several orders of magnitude \cite{makridakis1993accuracy}. As such, we optimize our QRF models for the root-mean-squared-logarithmic-error (RMSLE) of the data to best capture logarithmic-scale differences in the target (DFT calculated CBS energy corrections). The RMSLE is given by several definitions. One of them, given in Ref. \citeonline{sutton2019crowd} and employed in the SciKit-Learn package that is widely used by machine-learning practitioners~\cite{scikit-learn}, reads:
\begin{equation}
\text{RMSLE+1} = \sqrt{\frac{1}{N} \sum_{i = 1}^{N} (\log (y_i(\vec{x_i})+1) - \log(h(\vec{x_i})+1))^2}.
\label{eq:rmsle}
\end{equation}

The $\vec{x_i}$ are the feature vectors (combinations of feature values such as $C^A*\Delta E^{A}(N_b)$). The $y_i(\vec{x_{i}})$ values refer to the DFT calculated CBS energy corrections which our model tries to predict. Other authors do not employ the addition of one in the logarithm argument in the root mean squared error (RMSE) and instead add a small, different $\epsilon$ value instead~\cite{jachner2007statistical}. The addition of a small value in the logarithm argument is done to avoid undefined logarithmic arguments of zero. Note that due to the way we defined $\Delta E^{AB}(N_b)$ in eq. \ref{eq:delta_learning}, the values of $\Delta E^{AB}(N_b)$ are always negative by the variational principle, which states that the total energy must stay the same or decrease when the basis-set size is increased~\cite{giustino2014materials}. To employ the RMSLE as a metric, the targets (what our model tries to predict) should be positive valued. We satisfy this constraint by setting $y_{i}(\vec{x_i})$ equal to $-\Delta E^{AB}(N_b)$.

This use of $\epsilon = 1$, however, as in the RMSLE+1, is not ideal for targets much less than one, since in the Taylor expansion for $x << 1$ we have $log(x+1) \approx x$ and we arrive back at metric similar to the MAE that gives more weight to larger targets. Since our CBS energy convergence criteria is 1E-4 eV/atom we are motivated to use this value as our $\epsilon$. We term this metric RMSLE+1E-4.

\begin{equation}
\text{RMSLE+1E-4} = \sqrt{\frac{1}{N} \sum_{i = 1}^{N} (\log (y_i(\vec{x_i})+10^{-4}) - \log(h(\vec{x_i})+10^{-4}))^2}.
\label{eq:rmsle_plus_1E-4}
\end{equation}

Recall that QRF model in the combination model is trained on the stoichiometric residuals. The stoichiometric model may under or overestimate the CBS corrections giving the residual positive and negative signs. As such we cannot use the RMSLE to optimize the combination model. Instead we turn to a different metric, namely the  the symmetric mean absolute percentage error (sMAPE)~\cite{makridakis1993accuracy}, which is another popular metric for targets that vary orders of magnitude and is defined as:
\begin{equation}
sMAPE = \sum_{i=1}^{N}  \frac{ \lvert h( \vec{x_{i}} ) - y_{i}(\vec{x_i}) \rvert }{\frac{1}{2} \lvert y_{i}(\vec{x_i}) \rvert + \frac{1}{2} \lvert h( \vec{x_{i}} )\rvert} \times 100.
\label{eq:smape}
\end{equation}
Note, the closely related mean absolute percentage error (MAPE) is defined with a different denominator containing only the target as:
\begin{equation}
MAPE = \sum_{i=1}^{N}  \frac{ \lvert h( \vec{x_{i}} ) - y_{i}(\vec{x_i}) \rvert }{\lvert h( \vec{x_{i}} )\rvert} \times 100.
\label{eq:mape}
\end{equation}

Note that the MAPE is unbounded from above while the sMAPE is at most 200\% when either the target or prediction is zero. This fact means the sMAPE metric avoids issues where the target ($y_i(\vec{x_{i}})$) is close to zero and causes the value of the MAPE metric to explode~\cite{makridakis1993accuracy}. For this reason we optimize our models for the sMAPE rather than the MAPE. The sMAPE is often employed to optimize machine learning models operating on time series data where the target can grow exponentially~\cite{jaganathan2020combination}.  We also consider, however, the MAPE as a metric in our results since it is easier to comprehend. We experimented with training our QRF models by minimizing the sMAPE and saw slightly worse performance on the training data set in terms of the RMSLE+1E-4 and sMAPE as compared to when minimizing for the RMSLE+1E-4. Besides the RMSLE+1, RMSLE+1E-4, sMAPE and MAPE, we also include the MAE for completeness and the 95\% quantile absolute error (95\% quantile metric for short). This last metric represents the 95\% quantile of the absolute errors (AE) for each model.

\section*{Results}

We analyze the performance of three models on the data, namely the stoichiometric model, the QRF model and the combination model. Table \ref{tab:rf_model_total_energy} summarizes the most relevant metrics obtained for the test and training data. Note that the models are trained separately on FHI-aims and \exciting\ data. Recall also that the stoichiometric model has no free parameters to learn via training on the data.

\begin{table}[bh!]
\centering
\begin{tabular}{| p{2.25cm} | p{2.2cm} | p{2.19cm} | p{2.1cm} | p{1.9cm} | p{1.9cm} | p{2.0cm} |}
\hline
 & \multicolumn{3}{c|}{\textbf{\texttt{exciting}}} &   \multicolumn{3}{c|}{FHI-aims} \\
\hline
Metric &  Stoichiometric & QRF & Combination & Stoichiometric & QRF & Combination \\

\hline
sMAPE (\%)  & 28.7 \:\:\: (29.6) & 24.2 \:\:\: (10.6) & 26.4 \:\:\: (13.3)  & 48.9 \:\:\: (53.3) & 13.9 \:\:\: (10.6) & 14.84 \:\: (10.8) \\
\hline
MAPE (\%) & 27.9 \:\:\: (30.6) & 27.3 \:\:\: (11.6) & 27.2 \:\:\: (27.8) & 58.2 \:\:\: (70.1) & 14.9 \:\:\: (29.0) & 21.49 \:\: (33.6)\\
\hline
RMSLE+1  & 0.247 \: (0.016) & 0.202 \: (0.069) & 0.187  \: ( 0.061) &  0.203 (0.189) & 0.040 (0.015) & 0.031 \:  (0.014)\\
\hline
RMSLE+1E-4  & 0.383 \: (0.376) & 0.318 \: (0.150) & NA &  0.663 (0.916) & 0.218 (0.213) & NA \\
\hline
MAE (eV/atom) & 1.556 \: (1.250) & 1.437 \: (0.313) & 1.360 \: (0.260) & 0.257 (0.212) & 0.036 (0.013) & 0.030   \:  (0.014)\\
\hline
95\% Quantile (eV/atom) & 5.900 \: (8.018) & 7.795 \: (1.584) & 8.472 \: (1.340) & 1.520 (1.420) & 0.256 (0.072) & 0.177  \:  (0.074)\\
\hline
\end{tabular}
\caption{Metrics for the QRF and the combination models for \exciting\ and FHI-aims on held-out test data. The stoichiometric model performance is shown for comparison. The 95\% quantile metric refers to the 95\% quantile of the absolute error between the model and the DFT CBS corrections. The corresponding training data metrics are shown in parentheses. Note that \exciting\ data contains larger calculated DFT $\Delta E^{AB}$ targets since the basis-set size was controlled manually whereas the FHI-aims basis variation stopped at the discrete \emph{minimal} option given by the code. This results in larger MAE and maximum error values for the \exciting\ models.}
\label{tab:rf_model_total_energy}
\end{table}

The QRF models perform better than the stoichiometric models for all metrics except the 95\% quantile of absolute errors where it does slightly worse for \exciting. In general the CBS energy corrections are larger for \exciting\ than for FHI-aims, since the LAPW nature allows for manually reducing the basis-set size to close to zero~\citeonline{carbogno2022numerical}. We see this fact in the standard deviation of the CBS energy corrections of the test data which is 8.127 and 0.738 eV/atom for \exciting\ and FHI-aims, respectively. The larger 95\% quantile of absolute errors indicates that the stoichiometric model does slightly better than the QRF when targeting very large CBS energy corrections on the order of several eV/atom.

\begin{figure}[ht]
\centering
\includegraphics[width=0.8\linewidth]{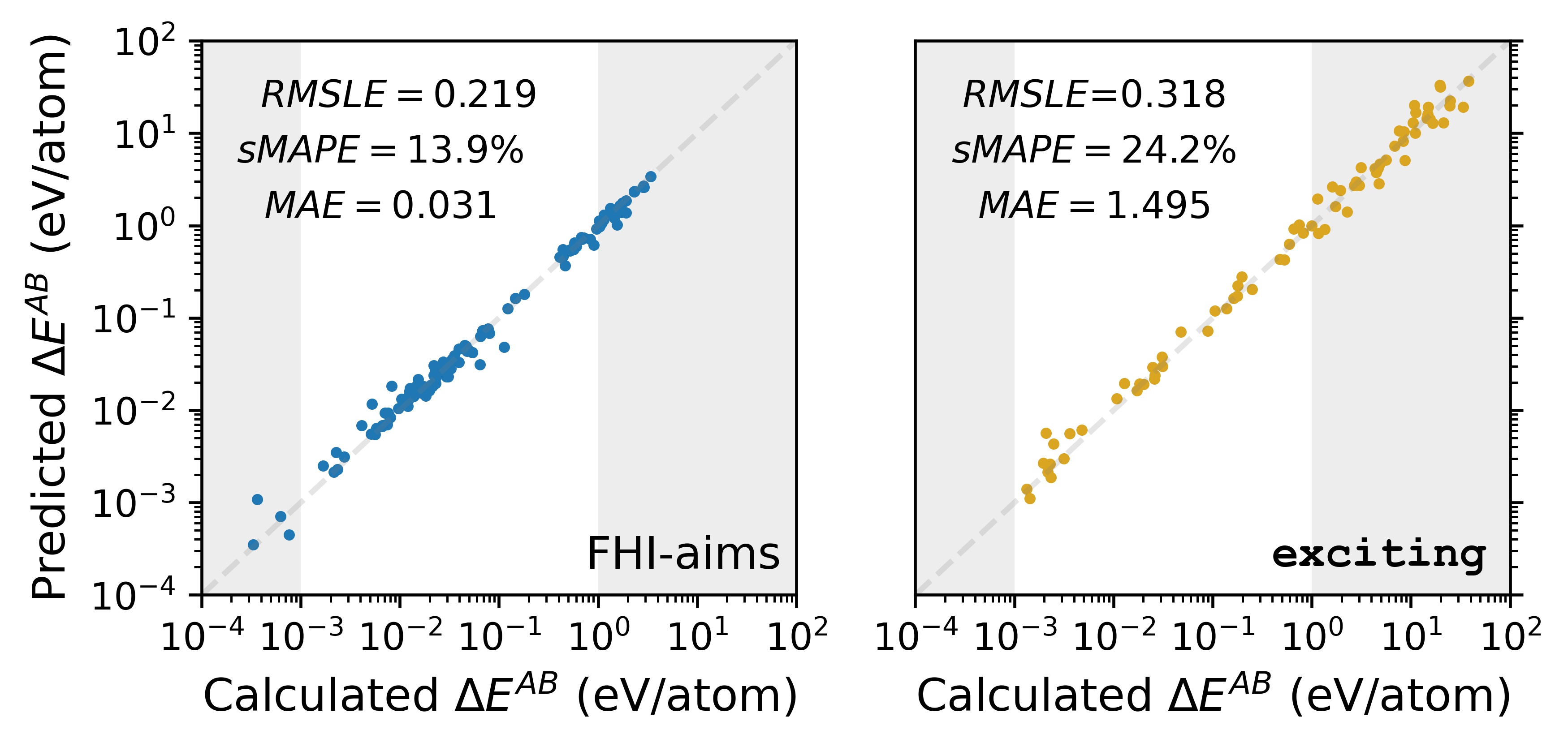}
\caption{Predictions of total-energy differences of the QRF model for FHI-aims (left) and \exciting\ (right) data, plotted against the respective DFT results. Relevant error metrics are added to help interpret the quality of the fit. Note the logarithmic axes. The region of DFT calculated $\Delta E^{AB}$ values between 1 meV/atom and 1 eV/atom is plotted with a lighter shade since these data are of particular interest to DFT practitioners. The RMSLE+1E-4 metric is labeled RMSLE.}
\label{fig:predicted_vs_DFT}
\end{figure}

The QRF models predictions are plotted against the DFT targets in fig. \ref{fig:predicted_vs_DFT}. Note that logarithmic scales are used to visually capture several orders of magnitude in the total-energy corrections. Relevant metrics, the sMAPE and MAE, are shown to help the reader understand the quality of fit. The RMSLE+1E-4 is also shown (labeled RMSLE) as the metric that is optimized during training.

For the MAPE metric in the FHI-aims case, surprisingly, the QRF model performs slightly better in the test data set than in the training data set. This may be due to the fact the QRF model is trained to fit the RMSLE+1E-4 metric and not the MAPE metric. We do not, in contrast, see a larger training error for the other five metrics of table \ref{tab:rf_model_total_energy}.

\begin{figure}[h]
\centering
\includegraphics[width=0.8\linewidth]{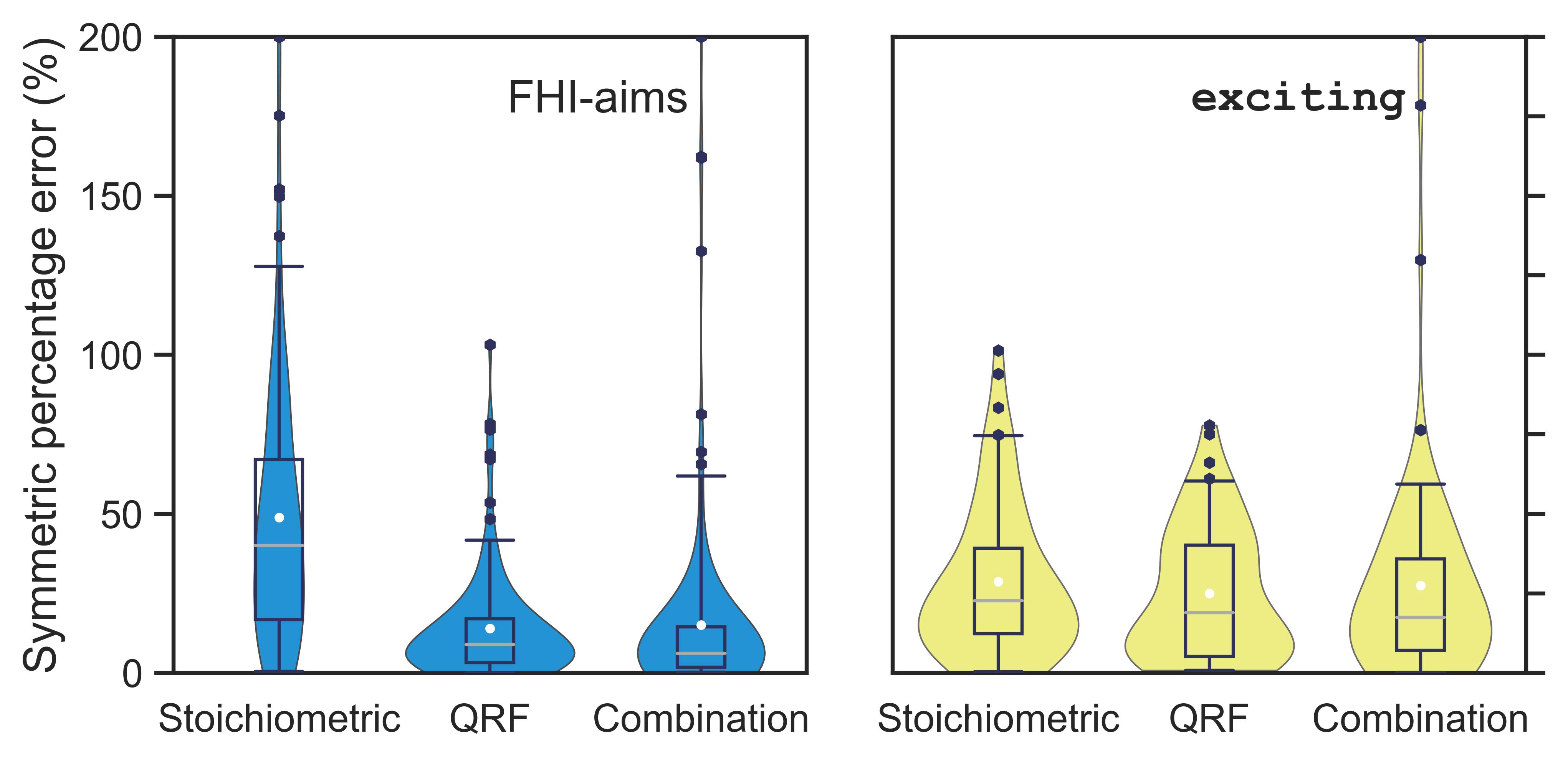}
\caption{Distribution of symmetric percentage errors of the stoichiometric, QRF, and combination model on held-out test data for FHI-aims (left) and \exciting\ (right) as violin plots. The box plot on top shows the 5\%, 25\%, 50\%, 75\% and 95\% quantiles. The white dot shows the mean symmetric percentage error. The black dots indicate values that fall outside of the 5\% and 95\% quantiles. We do not see the 5\% quantiles in the figure or data points that are smaller than the 5\% quantile since the 5\% quantiles are close to zero.}
\label{fig:violin_plot}
\end{figure}

We observe that the stoichiometric model for FHI-aims does quite poorly (test sMAPE of 48.9\%) whereas the stoichiometric model performs better on the \exciting\ data set with an sMAPE of 28.7\%. This is understood as the LAPW basis of \exciting\ is gradually improved by increasing the basis-set size parameter $RK_{\text{max}}$. The atomic-centered orbitals used in FHI-aims, on the other hand, give rise to a discrete basis-set size, specified as tiers ({\it minimal, tier1, tier2}) which correspond to more abrupt changes in basis-set quality. However, our non-linear QRF model provides a pretty good model for the CBS limit of FHI-aims. The discrete and piece-wise nature of its basis sets may explain why the very non-linear and piece-wise nature of the RF models succeeds on the FHI-aims data set.

The combination models perform worse than the QRF models in terms of test sMAPE and MAPE for both DFT codes. (Recall that the combination models are optimized for the sMAPE.) The combination models do, however, perform better in terms of RMSLE+1 and MAE. As discussed earlier, these metrics (with the 95\% quartile AE) generally favor larger targets in our data set closer to 1 eV/atom. The hypothesis that the combination model will outperform the QRF model appears false over the wide range of targets but true for very large targets. This may be due to the fact the residuals of the stoichiometric model are in general quite large, \eg the training sMAPE is 28.7\% and 48.9\% for \exciting\ and FHI-aims respectively, which make training a QRF on these residuals too difficult a task.

\begin{figure}[h]
\centering
\includegraphics[width=0.70\linewidth]{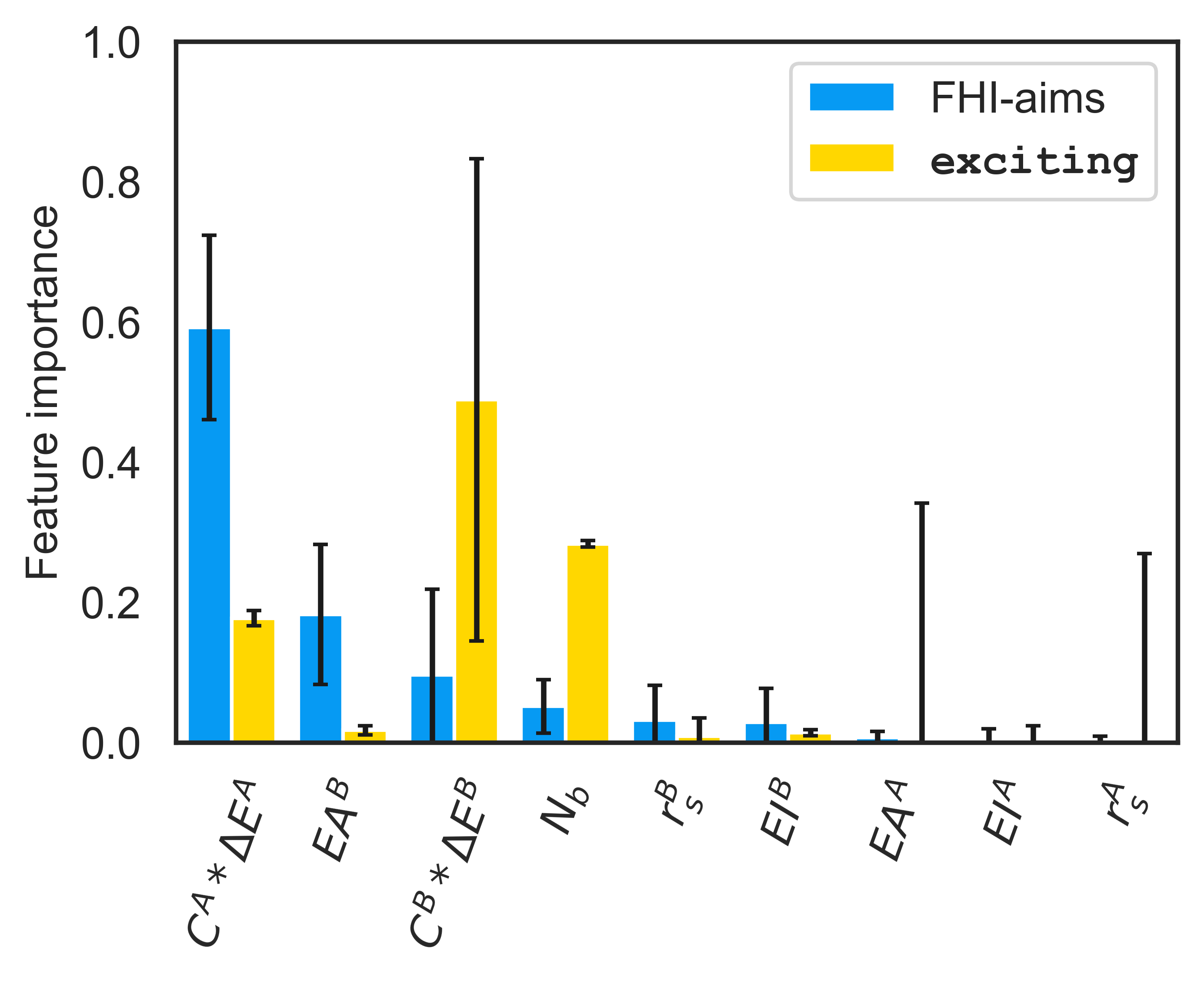}
\caption{Feature (Gini) importances of the QRF models for FHI-aims and \exciting. Black lines show the standard deviation in the feature importance across trees in the forests. The features are ordered in terms of their importance for the FHI-aims QRF model. $C^{A}*\Delta E^{A}$ and $C^{B}*\Delta E^{B}$ are the first and second terms in the stoichiometric model, respectively. $EA^{B}$ and $EI^{B}$, respectively, are the electron affinity and ionization potential of the elemental solid composed of element B, computed with FHI-aims. $r_{s}^{B}$ is the mean radius for the s-like pseudo orbital of elemental solid $B$ (computed with FHI-aims). $N_b$ is the basis-set size for the respective code.}
\label{fig:feature_importances_rf}
\end{figure}

The violin plots in fig. \ref{fig:violin_plot} display the distributions of symmetric percentage errors (SPE) on the test data for all models. Violin plots \cite{hintze1998violin} combine kernel-density plots with box plots, where the latter shows the model's SPE quantiles (5\%, 25\%, 50\%, 75\%, 95\%). Outliers are plotted with dots above the respective 95\% level. Note, the 95\% quantile of SPE is not the same as the 95\% quantile of AE metric in table \ref{tab:rf_model_total_energy}. The kernel-density plots, underneath the box plots, provide estimates for the probability-density for the SPE, \eg they estimate the likelihood of the prediction errors in a given range when using the model. For FHI-aims, the QRF model concentrates the SPE around 10\% while the stoichiometric model shows a corresponding thin distribution spanning a much larger range of SPE. The FHI-aims combination model appears to share many of the very large errors of the stoichiometric model. This is expected since the former employs a QRF that is fit on the residuals of the latter. When these residuals are sporadically very large, as is the case for both codes containing SPEs from their respective stoichiometric models of close to 100\%, the combination model has a difficult task to fit these residuals.

The \exciting\ QRF model has a smaller median 25\% quartile and 95\% quantile than the stoichiometric model but the 75\% quartile of the QRF model is larger. The larger 95\% quantile of the stoichiometric model is likely the reason why  the stoichiometric model returns a significantly worse sMAPE. This means that the QRF model has less very large SPE which is a desired behavior. The combination model for \exciting\ shows the best 75\% and 95\% quantile SPE but the worst outlier behavior with several data points found between 100\% and 200\%. This is likely a similar effect as seen in the FHI-aims combination model. No meaningful trends can be identified for the  QRF models for the predictions on test data of both codes that returned SPE above the 95\% quantile.

The Gini importance, or mean decrease of impurity (MDI)~\cite{breiman2001random}, of the features fed into the QRF models for both codes is displayed in fig. \ref{fig:feature_importances_rf}. The figure quantifies, for each variable, the QRF's ability to reduce the optimization metric using splits on that feature. Both codes' QRF models strongly depend primarily on either $\Delta E^{A}(N_b)$ or $\Delta E^{B}(N_b)$, which are the two features of the stoichiometric model. That these variables are effective in describing the CBS energy correction of the binary compounds comes as no surprise considering the overall success of the stoichiometric model~\cite{carbogno2022numerical}. $EA^{B}$, the electron affinity of the less electronegative element, \emph{B}, in the binary, is the second and fourth most important feature for the FHI-aims and \exciting\ QRF models.

The basis-set size variable, $N_b$, turns out to be the second and fourth most selected feature for \exciting\ and FHI-aims respectively. For FHI-aims, the basis-set size variable, $\Delta SB^{AB}_{PVE}$, is a single scalar. It maps an $s$-like orbital and a $d$-like orbital with the same weight. This loss of information in the mapping may be why we do not see the feature playing an important role. We experimented with feeding the model additional basis-set size variables, namely, the numerical and basis-set size settings encoded as integers but saw no improvement in CV performance. The fact that the features $\Delta E^{A}(N_b)$ and $\Delta E^{B}(N_b)$ are functions of the basis-set size may explain why the basis-set size variables themselves are not important since it is implicitly included in these variables.

The improvement of the QRF models over the corresponding stoichiometric models comes from two sources, the ability to use more variables and the ability to express non-linear functions. Based on the non-negligible feature importance of the basis-set size variable and atomic chemical data such as the electron affinity, we learn that the added features are important. However, this is not the only source of improvement. To provide evidence for it, we trained a linear model ($\ell$-1 and $\ell$-2 regularization) with these additional variables and found no significant improvement over the stoichiometric model. Therefore, we conclude that the QRF model benefits also from its non-linear nature.

\begin{figure}[h]
\centering
\includegraphics[width=0.8\linewidth]{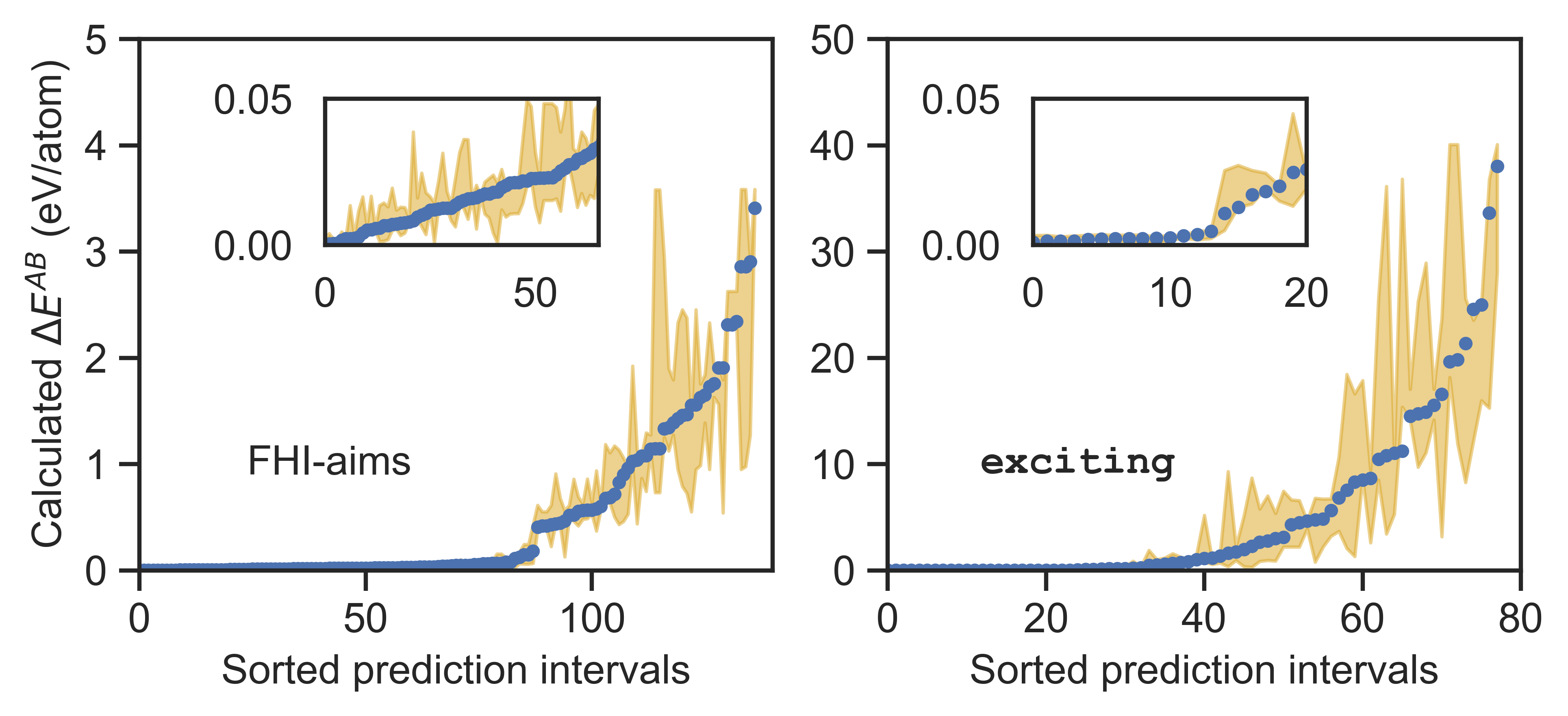}
\caption{Prediction intervals (95\%) for the QRF models for FHI-aims (left) and \exciting\ (right). The blue points indicate the calculated $\Delta E^{AB}$ DFT values. The yellow shaded areas show the prediction intervals for each DFT value target (blue point). The data (DFT values and associated prediction intervals) are ordered from left to right in terms of increasing DFT values. The insets zoom into the respective region of smaller values. Note, the x-axes are different for both codes since there is varying amount of data for each code due to the different basis-set size parameters. As a result, the y-axes are also different. \exciting\ contains larger calculated DFT $\Delta E^{AB}$ values since the basis-set size was controlled manually whereas the FHI-aims code basis variation stopped at the discrete \emph{minimal} option given by the code.}
\label{fig:prediction_intervals_rf}
\end{figure}

QRF models not only provide predictions but also associated prediction intervals based on the training data. Prediction intervals provide the user with an estimate for the uncertainty of the QRF prediction~\cite{heskes1996practical}.  When introducing quantile random forests, Meinshausen~\cite{meinshausen2006quantile} uses 95\% prediction intervals which are the difference between the 2.5\% quantile and 97.5\% quantile. The target CBS energy corrections and their associated 95\% prediction intervals are shown in fig. \ref{fig:prediction_intervals_rf}. They contain 92.7\% of the held-out test data for FHI-aims and 85.9\% for \exciting\. If the estimated 2.5\% and 97.5\% quantiles from our model were correct, we would expect 95\% of the data to fall in this range. In other words, on average 2.5\% of all unseen data points from the test data set should be above and below the 95\% prediction interval. This implies that our prediction intervals for both best performing models cover slightly less of the test data than desired. That said, we find the deviation acceptable with only 10 and 11 data points for FHI-aims and \exciting, respectively, being outliers. We note in passing that 99\% prediction intervals (or other values) can also be easily created using the accompanying Jupyter Notebook if the user desires more conservative uncertainty estimates.

\begin{figure}[ht]
\centering
\includegraphics[width=0.8\linewidth]{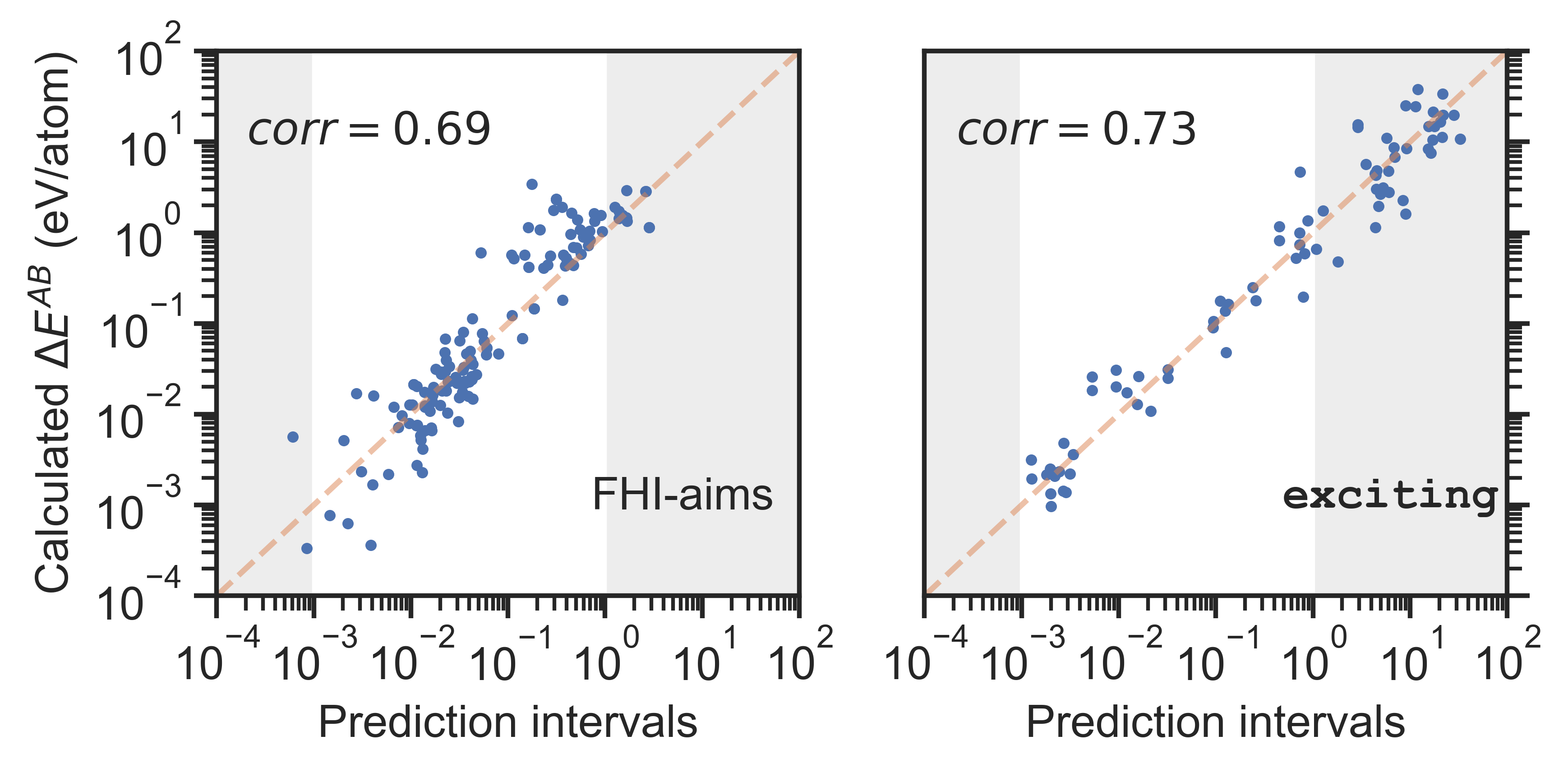}
\caption{The QRF model's prediction interval size versus DFT CBS energy correction targets for FHI-aims (left) and \exciting\ (right) data. Note, the logarithmic axes. This plot shows the effect of the calculated DFT targets on the associated prediction interval sizes of the QRF model. Larger/smaller $\Delta E^{AB}$ targets are generally associated with larger/smaller prediction intervals. The region of DFT calculated $\Delta E^{AB}$ values between 1 meV/atom and 1 eV/atom are plotted with a lighter shade since these data are of particular interest to DFT practitioners.}
\label{fig:prediction_intervals_vs_true_targets}
\end{figure}

We further analyze the 95\% prediction intervals by plotting them against the calculated DFT values (the model target) in fig. \ref{fig:prediction_intervals_vs_true_targets}. In general, we would expect larger calculated $\Delta E^{AB}$ values to have larger associated model prediction intervals. This is indeed what we see. The \exciting\ prediction intervals have a Pearson correlation of 0.73 with the calculated $\Delta E^{AB}$ values. For FHI-aims, the correlation of the prediction intervals is lower although still significantly positive at 0.69.

\section*{Discussion}

The QRF models allow us to perform CBS extrapolation of the total energy per atom. In this case, $E^{AB}(N_b)$, is known, and the model estimates the correction, $\Delta E^{AB}(N_b)$, the sum of both giving us the CBS estimate, $E^{AB}(N_\infty)$ (eq. \ref{eq:delta_learning}). Assessing the overall performance of our models, for both codes, we find that the QRF models achieve improved results in all metrics except for the 95\% quantile AE for which the \exciting\ model performs slightly worse. This indicates that for the largest CBS extrapolation corrections of the \exciting\ data set, which are around 20-40 eV/atom the stoichiometric model is preferred. This knowledge gives us a better understanding of the QRF model's domain of applicability --\ie the user should not expect more reliable CBS corrections than this from the stoichiometric model for calculated corrections of over 20 eV/atom. Overall, we understand that the improvement of the QRF models over the stoichiometric models comes from the addition of new features and the ability to fit non-linear distributions.
We also find the combination models to perform worse than the QRF models in terms of our metrics that cover a wide range of targets (sMAPE, MAPE) and as such we don't recommend the use of the combination model.

The QRF models also provide prediction intervals which offer the user a quantitative estimate of how uncertain the model is in each prediction it makes. The 95\% prediction intervals are positively correlated with over 0.69 Pearson correlation for both codes. We expect the model to have larger prediction intervals for larger targets and we see this with the 95\% prediction intervals which have over 0.69 Pearson correlation for both codes. The 95\% prediction intervals contain slightly less than 95\% of the held-out calculated CBS energy corrections. In general, however, the correlation of the prediction intervals and the percent of data they cover indicate the prediction intervals of the QRF models to be well-behaved.

The QRF models offer users of materials databases a quantitative assessment of how far a total-energy result is from the CBS limit, which helps to evaluate how these results can be reused/repurposed. Looking forward, we believe that this work will allow such databases to provide estimated CBS corrections for hosted data. We find the overall performance of the QRF models in terms of sMAPE on the test data set of less than 25\% for \exciting\ and less than 15\% for FHI-aims acceptable, especially considering that we provide prediction intervals that indicate the precision of the estimate (and thereby the domain of applicability of the model). This will help non-experts unlock the large potential these databases have in many fields (\eg medical, transportation, energy). For instance, data that were simulated for a molecular dynamics investigation,  likely have a large CBS correction and might be unsuitable for investigations where high precision is required. On the opposite case, data performed with very high precision settings, even coming from different sources, will have a low CBS correction and might be suitable for a wide range of machine-learning tasks. We also believe that this work may have future applications in recommending a basis-set size for DFT practitioners for achieving a certain degree of precision before a calculation is performed and thereby saving computational expenses.

We expect highly expressive non-linear models such as neural networks and related methods to offer an opportunity to improve CBS-limit predictions for DFT data~\cite{speckhard2023neural}. However, these models are notoriously data-hungry, and more data are needed before they can be used effectively. The authors plan high-throughput calculations to obtain more dedicated data for a systematic analysis. This not only includes total energies but also tackling more involved properties such as electronic or elastic properties, and more.

\section*{Data Availability}
The raw DFT data, \ie input and output files for both \exciting\ and FHI-aims are hosted in the \href{https://repository.nomad-coe.eu}{NOMAD Repository}, under the following DOIs:
\exciting: \href{https://dx.doi.org/10.17172/NOMAD/2020.07.15-1}{DOI:10.17172/NOMAD/2020.07.15-1},
FHI-aims: \href{https://dx.doi.org/10.17172/NOMAD/2020.07.27-1}{DOI:10.17172/NOMAD/2020.07.27-1}.

\section*{Code Availability}

The code used to generate all figures and train all models in this paper can be found on the NOMAD AI-Toolkit~\cite{sbailo2022nomad} at this link: \url{https://nomad-lab.eu/aitutorials/error-estimates-qrf}.

\bibliography{main}

\section*{Acknowledgements}
This work received partial funding by the by the IMPRS {\it Elemental Processes in Physical Chemistry}, the German Research Foundation (DFG) through the CRC 1404 (FONDA), Projektnummer 414984028, and the Max-Planck Graduate Center for Quantum Materials. The NOMAD Center of Excellence, grant agreement Nº 951786 under the Union’s Horizon 2020 research and innovation program is appreciated. The hyper-parameter search for the QRF models in this work were conducted with computing resources under the project bep00098.

\section*{Author Contributions Statement}

D.S. performed the data analysis and wrote the quantile-random-forest analysis software. C.C. advised on the machine learning and the FHI-aims calculations. L.G. advised on the machine learning. S.L. gave advice with respect to the \exciting\ calculations. M.S. and C.D. designed the research questions and advised on all matters of the research. All authors contributed to the manuscript.

\section*{Competing interests} The authors have no competing interests.

\section*{Supplementary Information}

\begin{figure}[ht]
\centering
\includegraphics[width=0.8\linewidth]{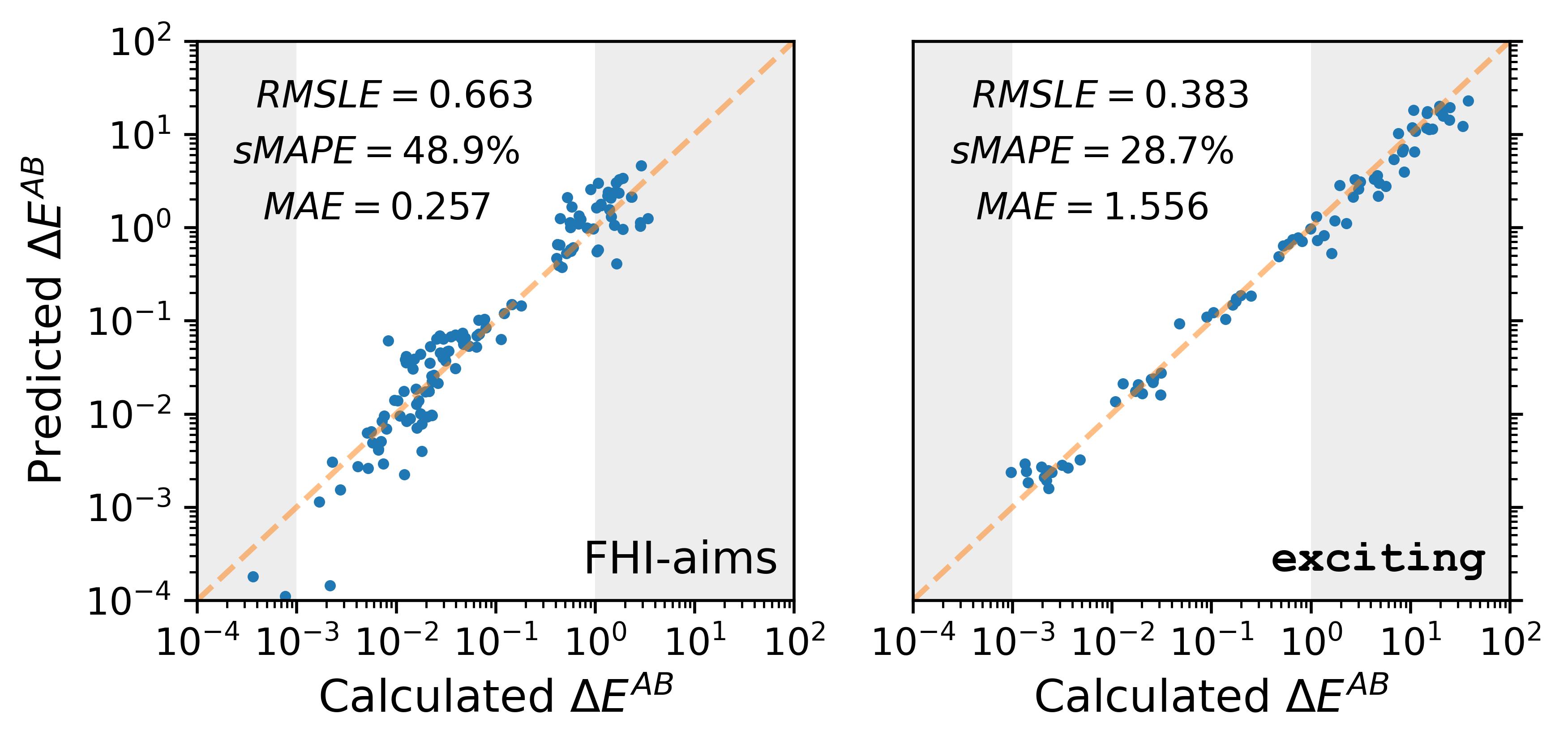}
\caption{Stoichiometric model predictions of $\Delta E^{AB}$ for  FHI-aims (left) and \exciting\ (right). The regions of particular interest to DFT practitioners, \ie calculated values between 1 meV/atom and 1 eV/atom, are highlighted by the white background. The RMSLE+1E-4 metric is labeled RMSLE.}
\label{fig:stoichiometric_predicted_vs_DFT}
\end{figure}

The stoichiometric models predictions on the test data set are seen in fig. \ref{fig:stoichiometric_predicted_vs_DFT}.

\subsection*{Decision Tree Optimization of RMSLE}
\label{sec:DT_RMSLE}
Quantile random forests are built on decision trees which typically are piece-wise constant functions, defined as 
\begin{equation}
 h(x_i) =  \sum \limits _{m=1}^{M} c_{m}I \: (x_i \in R_{m}) .
  \label{eq:piecewise}
\end{equation}
The index \emph{i} refers to an individual input in the data set identifying the compound, \emph{AB} and basis-set size, $N_b$. The indicator function $I$ is equal to one if the input data point $x_i$ belongs to a region of the input data space $R_m$. The index \emph{m} refers to the different regions of input space which the decision tree partitions. These trees estimate a constant $c_m$ when the input data point $x_m$ falls into a region $R_m$, where $m$ runs over the entire data set. Note in general, and in the models used in this paper, the input features, $x$, are multidimensional (\eg more than one one-dimensional feature is used for inference). The constants are chosen to minimize the metric of choice, in our case the RMSLE, for each region, $R_m$ of input space. We minimize the mean squared logarithmic error (MSLE) for simplicity since it is equivalent to minimizing the RMSLE since the square root function is convex. Our loss (cost) function, labeled $J$, is then defined as in eq. \ref{eq:cost_function}.
\begin{equation}
 J = \sum \limits _{i=1}^{i=N} (log(y(x_i)+\epsilon) - log (\sum \limits _{m=1} ^{M} c_{m} I \: (x_{i} \in R_{m}))+\epsilon)^2
\label{eq:cost_function}
\end{equation}

We take the derivative of the cost in this region with respect to the constant, $c_{m'}$, and set this derivative equal to zero. The constant, $c_{m'}$ for a region $R_{m'}$ is determined only by the input data that falls in that region. The model in eq. \ref{eq:piecewise} predicts the constant $c_{m'}$ for data that falls in that region $R_{m'}$. We compare that prediction with the true value $y(x_i)$. To simplify our analysis, we focus on a single region $R_{m'}$ labeled with integer $m'$ and set out to find the value of the constant $c_{m'}$ that minimizes the loss of that region, $J(c_m')$. The loss function thus becomes
\begin{equation}
 J(c_{m'}) = \sum \limits _{x_i \in R_{m'}} (log(y(x_i)+\epsilon) - log (c_{m'} +\epsilon))^2 .
\label{eq:cost_function_region_prime}
\end{equation}
In this region, the only data points that affect the loss satisfy the condition $x \in R_{m'}$. Moreover, in this region we can replace our sum of indicator functions (piece-wise constant tree) defined in eq. \ref{eq:piecewise} with a single constant, $c_{m'}$ for this region. We then arrive at eq. \ref{eq:cost_function_derivative_zero}.
\begin{equation}
 \frac{dJ(c_{m'})}{dc_m'} = -2 \frac{1}{(c_{m'}+\epsilon)}\sum \limits _{x_i \in R_{m'}} (log(y(x_i)+\epsilon) - log (c_{m'} +\epsilon)) = 0 .
\label{eq:cost_function_derivative_zero}
\end{equation}
We note immediately that $c_{m'}=-1$, leaves our loss function gradient with a logarithmic argument of zero and a division by zero which is undefined and not a solution we desire. If $c_{m'}\neq-1$, we arrive at
\begin{equation}
\sum \limits _{x_i \in R_{m'}} (log(y(x_i)+\epsilon) = N_{m'}log (c_{m'} + \epsilon))
\label{eq:cost_function_derivative_zero_soln_two}
\end{equation}
where $N_{m'}$ is the number of data points in the region (\ie satisfying $x_m \in R_{m'}$). The use of the RMSLE as a metric leads to 
\begin{equation}
c_{m'} = \exp \big( -N_{m'} \: \sum \limits _{x_i \in R_{m'}} (log(y(x_i)+\epsilon)\big) - \epsilon
\label{eq:cost_function_derivative_zero_soln_two_final}
\end{equation}
as the choice of optimal constants for each region in our decision tree. Note, if we had chosen the MSE as a metric for the above analysis, we would have found the following constants \ie simply the average of targets in the region.
\begin{equation}
c_{m'} = \frac{1}{N} \sum \limits _{x_i \in R_{m'}} y(x_i)
\label{eq:constant_mse}
\end{equation}

\end{document}